\documentstyle[preprint,prb,aps]{revtex}
\begin{document}
\draft
\title{Interaction-Induced Enhancement of Spin-Orbit Coupling in Two-Dimensional Electronic System}

\author{Guang-Hong Chen and M.~E.~Raikh}
\address{Department of Physics, University of Utah, 
Salt Lake City, Utah  84112}

\maketitle
\begin{abstract}
\baselineskip=0.90cm
We study theoretically the renormalization of the spin-orbit coupling constant of two-dimensional electrons by electron-electron interactions. We demonstrate that, similarly
 to the $g$ factor, the renormalization corresponds to the
 enhancement, although the magnitude of the enhancement is
 weaker than that for the $g$ factor. For high electron concentrations
 (small interaction parameter $r_s$) the enhancement  factor is
 evaluated  analytically within the static random phase approximation.
 For large $r_s\sim 10$ we use an approximate expression for effective
 electron-electron interaction, which takes into account the local field factor,
 and calculate the enhancement  numerically. We also study the interplay between the interaction-enhanced Zeeman splitting and interaction-enhanced spin-orbit coupling.
 \end{abstract}


\section{Introduction}
Early experimental studies of magnetotransport 
in two-dimensional (2D) electron
systems\cite{fang} indicated that the $g$ factor of electrons
 in these systems may differ significantly from its bulk value.
It was established\cite{fang} that the magnitude  of the $g$ factor
for electrons confined to $(100)$ Si surfaces exceeds  $g=2$
and   increases from $g=2.47$ to $g=3.25$ with  decreasing
the concentration of electrons from $6 \times 10^{12}$cm$^{-2}$ to 
 $10^{12}$cm$^{-2}$. 

Shortly after the publication of experimental results\cite{fang}
it was suggested by Janak\cite{janak} that the enhancement of the $g$ factor
can be accounted for by the electron-electron interactions.
The argument of Janak represents a 2D version of the Fermi  liquid theory\cite{pines}
and goes as follows. 
In applied weak magnetic field, $B$, the quasiparticle
energies for the
two spin projections can be written as
\begin{eqnarray}
\label{Janak}
E_{\uparrow}(k)=E^{(0)}(k)+\frac{\Delta_Z}{2} +\Sigma_{\uparrow}(k,E_{\uparrow}(k)), \nonumber \\
E_{\downarrow}(k)=E^{(0)}(k)-\frac{\Delta_Z}{2}+\Sigma_{\downarrow}(k,E_{\downarrow}(k)),
\end{eqnarray}
where ${\bf k}$ is the momentum, $E^{(0)}(k)=\hbar^2k^2/2m$ is the spectrum of a free electron,
$\Delta_Z=g\mu_BB$ is the bare Zeeman splitting, and $\Sigma(k,E_k)$ is the self-energy
\begin{equation}
\label{sigma}
\Sigma_{\uparrow,\downarrow}(k)=-\int\frac{d^2k^{\prime}}{(2\pi)^2}
V_{eff}(|{\bf k}-{\bf k}^{\prime}|)f_0(E_F-E_{\uparrow,\downarrow}(k^{\prime})) ,
\end{equation}
where $V_{eff}(q)$ is the Fourier component of the effective
interaction between the electrons, and $f_0$ is the Fermi function.
Solving the system Eq. (\ref{Janak}),  Eq. (\ref{sigma}) in the zero-temperature
limit, the effective $g$-factor can be presented as
\begin{equation}
\label{star}
g^{\ast}=\frac{\Delta_Z^{\ast}}{\mu_BB},
\end{equation}
where
\begin{equation}
\label{effective}
\Delta_Z^{\ast}=E_{\uparrow}(k_F)-E_{\downarrow}(k_F)=
\frac{\Delta_Z}{1-\frac{m^{\ast}}{m}\lambda_Z}.
\end{equation}
In Eq. (\ref{effective}) $m^{\ast}$ is the effective mass
\begin{equation}
\label{mass}
m^{\ast}=\hbar^2k_F\Biggl(\frac{\partial E_k}{\partial k}\Biggr)_{k_F}^{-1},
\end{equation}
and $k_F$ is the Fermi momentum.
The enhancement factor $\lambda_Z$ is given by
\begin{equation}
\label{lambda}
\lambda_Z=\frac{m}{(2\pi\hbar)^2}
\int_0^{2\pi}d\phi V_{eff}\Bigl(2k_F\sin\frac{\phi}{2}\Bigr) .
\end{equation}
In the random phase approximation (RPA) one has\cite{stern}
\begin{equation}
\label{diel}
V_{eff}(q)=\frac{2\pi e^2}{\varepsilon_0(q+\sqrt{2}r_sk_F)} ,
\end{equation}
for $q<2k_F$, where $\varepsilon_0$ is the dielectric constant 
of the material,  and
 $r_s=\sqrt{2}me^2/\varepsilon_0\hbar^2k_F$ is the interaction 
parameter of the 2D gas.
With $V_{eff}(q)$ in the form Eq. (\ref{diel}), 
$m^{\ast}$ and $\lambda_Z$ can be evaluated analytically yielding\cite{janak,yarl} 
\begin{equation}
\label{renorm'}
\lambda_Z={\cal F}(r_s),
\end{equation}
\begin{equation}
\label{renorm}
\frac{m}{m^{\ast}}=1-\frac{\sqrt{2}}{\pi}r_s+\frac{r_s^2}{2}+(1-r_s^2){\cal F}(r_s), 
\end{equation}
where the function ${\cal F}(r_s)$ is defined as
\begin{eqnarray}
\label{rpa1}
{\cal F}(r_s)=\frac{r_s}{\pi\sqrt{2-r_s^2}}\cosh^{-1}\Biggl(\frac{\sqrt{2}}{r_s}\Biggr),\hspace{1.2cm}
 r_s\leq \sqrt{2},\nonumber \\
{\cal F}(r_s)=
\frac{r_s}{\pi\sqrt{r_s^2-2}}\cos^{-1}\Biggl(\frac{\sqrt{2}}{r_s}\Biggr),\hspace{1.2cm}
 r_s\geq \sqrt{2}.
\end{eqnarray}
In the high-density limit ($r_s\ll 1$) the enhancement 
factor (\ref{renorm'}) takes the form (see also Ref. \onlinecite{ale})
\begin{equation}
\label{limit}
\lambda_Z=\frac{r_s}{\sqrt{2}\pi}\ln\Biggl(\frac{2^{3/2}}{r_s}\Biggr).
\end{equation}
Note that the theory\cite{janak} neglects the frequency dependence of $V_{eff}$.
As a result, Eq. (\ref{renorm}) predicts that interactions reduce the
effective mass. In fact, taking the frequency dependence into account\cite{lee}
leads to $m^{\ast}/m >1$ already within the RPA (see, however, the recent 
numerical simulations\cite{kwon}).

Later magnetotransport experiments\cite{luo,das} on quantum well structures
in narrow band semiconductors provided an evidence for
a splitting of the conduction band in a zero magnetic field.
The analysis of the beating patterns in electron 
Shubnikov-de Haas oscillations led the authors\cite{luo,das}
to the conclusion that such a splitting can be accounted for
by adding the spin-orbit (SO) term
\begin{equation}
\label{hso}
\hat{H}_{SO}=\alpha{\bf k}\cdot({\bf\bbox{ \sigma}}\times {\bf n}),
\end{equation}
to the Hamiltonian of a free electron.
Here  $\alpha$ is the SO coupling constant,  ${\bf k}$ is the wave vector,  ${\bf n}$ is the unit vector normal to the    plane of the
quantum well,
  $\bbox{\sigma}=(\sigma_{1},\sigma_{2},\sigma_{3})$  are the Pauli matrices. The term Eq. (\ref{hso}) was first introduced by 
Bychkov and Rashba\cite{rashba,rashba'} 
to explain the experimental results
on electron spin resonance\cite{klitz} and a cyclotron resonance 
of holes\cite{stor} in GaAs/AlGaAs heterostructures.

In order to obtain more
 detailed information  about the SO-induced splitting 
of the conduction band, the evolution of the Shubnikov-de Haas
oscillations with a tilting of magnetic field was traced\cite{luo'}. 
Subsequently, the energy spectrum of a 2D electron in
a tilted magnetic field in the presence of the SO coupling
was studied theoretically\cite{rash'',das1}. 

Recently a zero-field splitting in different 2D systems 
was inferred experimentally either from the Shubnikov-de Haas\cite{engels,ram,nitta,heida}
or from the commensurability oscillations\cite{shay} (in a spatially
modulated sample) patterns. 

In the domain of weak magnetic fields 
  the Shubnikov-de Haas oscillations are smeared out. However,
it was demonstrated both experimentally\cite{dress}
 and theoretically\cite{mathur} that the SO coupling
still manifests itself in this domain through the weak
localization corrections to the conductance.
Early works\cite{dress,mathur} in this direction
were succeeded by detailed studies\cite{lyanda}.

In the present paper we investigate theoretically the interplay
between the SO coupling and the electron-electron interactions.
Namely, we address the question whether the interactions
cause the renormalization of the coupling constant $\alpha$
in Eq. (\ref{hso}) as it is the case for the $g$-factor.

\section{Renormalization of the SO splitting}
We assume that the bare SO coupling is weak enough $\alpha k_F\ll E_F$. With the SO term Eq. (\ref{hso}) the Hamiltonian of non-interacting electrons can be presented 
in the form
\begin{equation}
\label{h0}
\hat{H}=E^{(0)}_{+}(k)\hat{P}^{+}({\bf k})+E^{(0)}_{-}(k)\hat{P}^{-}({\bf k}),
\end{equation}
where the projection operators $\hat{P}^{+}({\bf k})$ and $\hat{P}^{-}({\bf k})$ 
are defined as
\begin{equation}
\label{p+-}
\hat{P}^{+}({\bf k})=\frac{1}{2}\left(\begin{array}{cc}1 & ie^{-i\phi_{{\bf k}}}\\ -ie^{i\phi_{\bf k}} & 1\end{array}\right),\hspace{1.5cm}\hat{P}^{-}({\bf k})=1-\hat{P}^{+}({\bf k}),
\end{equation}
so that $\hat{P}^{+}({\bf k})\hat{P}^{-}({\bf k})=0$. In Eq. (\ref{p+-}) 
$\phi_{\bf k}=\arctan(k_y/k_x)$ is the azimuthal angle of the wave vector ${\bf k}$. The energy spectrum consists of two branches
\begin{equation}
\label{e0k}
E^{(0)}_{\pm}(k)=\frac{\hbar^2k^2}{2m}\pm\alpha k.
\end{equation}
Following the Fermi liquid theory, the self-energy in the presence of the SO 
coupling becomes an operator
\begin{equation}
\label{sigmak}
\hat{\Sigma}({\bf k})=\int\frac{d^2{\bf k}^{'}}{(2\pi)^2}V_{eff}(|{\bf k}-{\bf k}^{'}|)\biggl[\hat{P}^{+}({\bf k}^{'})f_0(E_F-E_{+}(k^{'}))+\hat{P}^{-}({\bf k}^{'})f_0(E_F-E_{-}(k^{'}))\biggr].
\end{equation}
Our main observation is that in the presence of the interaction, the operator $\hat{\Sigma}({\bf k})$ still retains the structure of Eq. (\ref{h0})
\begin{equation}
\label{form1} 
\hat{\Sigma}({\bf k})=\Sigma^+(k)\hat{P}^{+}({\bf k})+\Sigma^-(k)\hat{P}^{-}({\bf k}),
\end{equation}
where $\Sigma^{\pm}(k)$ are the {\em scalar} functions of the absolute value of the wave vector ${\bf k}$
\begin{eqnarray}
\label{sigmapmk}
\Sigma^{\pm}(k)=-\frac{1}{2}\int\frac{d^2{\bf k}^{'}}{(2\pi)^2}V_{eff}(|{\bf k}-{\bf k}^{'}|)\biggl[f_0(E_F-E_{+}(k^{'}))+f_0(E_F-E_{-}(k^{'}))\biggr]\\ \nonumber
\pm\frac{1}{2}\int\frac{d^2{\bf k}^{'}}{(2\pi)^2}\cos(\phi_{\bf k}-\phi_{{\bf k}^{'}})V_{eff}(|{\bf k}-{\bf k}^{'}|)\biggl[f_0(E_F-E_{+}(k^{'}))-f_0(E_F-E_{-}(k^{'}))\biggr].
\end{eqnarray}
For renormalized energy spectrum, we have
\begin{equation}
\label{ek}
E_{+}(k)=E_{+}^{(0)}(k)+\Sigma^{+}(k),\hspace{1.0cm}E_{-}(k)=E_{-}^{(0)}(k)+\Sigma^{-}(k).
\end{equation}
By solving the system Eq. (\ref{sigmapmk}) and Eq. (\ref{ek}), we get the following result for the  renormalized SO splitting
\begin{equation}
\label{deltaso}
\Delta_{SO}^*=E_{+}(k_F)-E_{-}(k_F)=\frac{\Delta_{SO}}{1-\frac{m^*}{m}\lambda_{SO}},
\end{equation}
where $\Delta_{SO}=2\alpha k_F$ is the bare SO splitting and the 
 renormalization factor is determined as 
\begin{equation}
\label{lambdaso}
\lambda_{SO}=\frac{m}{(2\pi\hbar)^2}\int_0^{2\pi}d\phi\cos\phi V_{eff}(2k_F\sin\frac{\phi}{2}). 
\end{equation}
If $ V_{eff}$ does not depend on $\phi$ (when interactions are short-ranged due
{\em e.g.}
 to the presence of a gate electrode close to the 2D plane), then we have 
$\lambda_{SO}=0$. However, in general, the integral is positive. Thus, we conclude that the exchange interaction leads to the {\em enhancement} of the SO coupling. Within the random phase approximation when $V_{eff}$ has the form Eq. (\ref{diel}), the integral (\ref{lambdaso}) can be  calculated analytically and expressed through the function ${\cal F}(r_s)$ as follows
\begin{equation}
\label{lamso} 
\lambda_{SO}=\frac{r_s^2}{2}-\frac{\sqrt{2}r_s}{\pi}+(1-r_s^2){\cal F}(r_s).
\end{equation}
Comparison of the last expression with Eq. (\ref{renorm}) indicates that
$1+\lambda_{SO}=m/m^{\ast}$. In fact, this relation holds not only
for $V_{eff}(q)$ in the form (\ref{diel}),
 but for arbitrary effective interaction and represents a 2D version
of the corresponing relation in the Fermi liquid theory\cite{abr}.
To verify this relation, it is convenient to perform transformation to the real
space where the interaction has the form $\tilde V_{eff}(\rho)$. Then from
Eqs. (\ref{mass}) and (\ref{lambdaso}) it is
easy to check that
\begin{equation}
\label{real}
\lambda_{SO}=\frac{m}{m^{\ast}}-1=\frac{m}{\hbar^2}\int_0^{\infty}d\rho\rho{\cal J}_1^2(k_F\rho)
\tilde V_{eff}(\rho),
\end{equation}  
where ${\cal J}_1(x)$ is the Bessel function. Combining (\ref{deltaso}) with (\ref{real}), we get
\begin{equation}
\label{simple}
\frac{\Delta_{SO}^{\ast}}{\Delta_{SO}}=1+\lambda_{SO}.
\end{equation}
In Fig. 1 we plot both $\lambda_Z$ and $\lambda_{SO}$ as a function of interaction parameter $r_s$. It is seen that $\lambda_Z$ is much bigger than $\lambda_{SO}$,
 which has a maximum at $r_s=0.52$ and does not exceed $6$~percent. On the other hand, it is known that at large $r_s$ the random phase approximation overestimates the screening effect which, in turn, suppresses $\lambda_{SO}$.
To extend the Fermi liquid description to higher $r_s$, it is customary \cite{mahan} to modify the random phase  dielectric function \cite{stern} as follows
\begin{equation}
\label{dielectric}
\varepsilon(q)=\varepsilon_0\Biggl(1-\frac{v(q)\chi_0}{1+v(q)G(q)\chi_0}\Biggr),
\end{equation}
where $v(q)$=$2\pi e^2/\varepsilon_0q$ is the Fourier component of the Coulomb interaction and  $\chi_0$=$-m/\pi\hbar^2$ is the Lindhard susceptibility of the free electron gas. The factor $G(q)$ (local field correction) describes the reduction of the screening at large $q$ (small distances). For $G(q)=0$ we recover Eq. (\ref{diel}) for the effective interaction $V_{eff}(q)=v(q)/\varepsilon(q)$.

In later works\cite{yar,yarl,san,yarlag} 
 a different approximation for $V_{eff}(q)$  was put forward
\begin{equation}
\label{effpoten}
V_{eff}(q)=v(q)+v^2(q)[1-G(q)]^2\chi(q),
\end{equation}
where $\chi(q)$ is defined as
\begin{equation}
\label{response}
\chi(q)=\frac{\chi_0}{1-v(q)(1-G(q))\chi_0}.
\end{equation}
For  the local field correction $G(q)$ the authors\cite{yar,yarl,san}
adopted the following form
\begin{equation}
\label{localfield1}
G(q)=\frac{G_{\infty}q}{\sqrt{q^2+q^2_{1}(r_s)}},
\end{equation}
where $q_1(r_s)=2a(r_s)k_F$, and  
$G_{\infty}(r_s)$, $a(r_s)$ are the numerical factors. 
Eq. (\ref{effpoten}) is written neglecting the spin-fluctuation-induced vertex corrections.
 Combining Eqs. (\ref{effpoten}), (\ref{localfield1}) and (\ref{lambdaso}), we get the following expression for the enhancement factor of SO coupling
\begin{equation}
\label{grs}
\lambda_{SO}=\frac{r_s}{4\pi\sqrt{2}}\int_0^{2\pi}d\phi\cos\phi\frac{\sqrt{a^2+\sin^2\phi/2}\biggl[G_{\infty}r_s+\sqrt{2(a^2+\sin^2\phi/2)}\biggr]-G_{\infty}^2r_s\sin\phi/2}{(r_s+\sqrt{2}\sin\phi/2)(a^2+\sin^2\phi/2)-G_{\infty}r_s\sin\phi/2\sqrt{a^2+\sin^2\phi/2}}.
\end{equation}
In Fig. 2 we present the dependence $\lambda_{SO}(r_s)$ calculated numerically
within the region up to $r_s\sim 8 $. Following\cite{yar,yarl,san}, we took 
$G_{\infty}(r_s)$ from numerical calculations (at discrete values of $r_s$)
of the pair correlation
function\cite{jonson} and following Ref.\onlinecite{yar} assumed 
$a(r_s)\approx 1.5G_{\infty}$ within the entire domain.    It is seen that instead
of falling down (as in Fig. 1) $\lambda_{SO}(r_s)$
increases with $r_s$, when the local factor is included. Note, however, 
that approximately 
constant value for $a(r_s)$ was established only within a limited interval
$r_s \leq 3$ in Refs.\onlinecite{yarl,yar}.
We used the same value for calculation at higher $r_s$ in order to illustrate that the
enhancement factor can take appreciable values in this domain.

The alternative approach to the effective interaction in 2D electron
gas with $r_s\gg 1$ is described in Ref. \onlinecite{iwam}. In this paper the local field
factor in the conventional form (\ref{dielectric}) of $\varepsilon (q)$
was fitted in such a way that the static characteristics of the system,
calculated with this  $\varepsilon (q)$, are  consistent with
 the Monte Carlo results of Tanatar
and Ceperley\cite{tan}. According to Ref. \onlinecite{iwam}
 the local field correction has the form
\begin{equation}
\label{iwamoto}
G(q)=\frac{1}{2}\Biggl[\frac{q}{(q^2+4b_1^2k_F^2)^{1/2}}+\frac{q}{(q^2+4b_2^2k_F^2)^{1/2}}\Biggr],
\end{equation}
where  the parameters $b_1(r_s)$, $b_2(r_s)$ are listed in Ref. \onlinecite{iwam}
at discrete values of $r_s$ up
 to $r_s=40$. The numerical results for $\lambda_{SO}$
calculated for these values by substituting  $G(q)$ in the form (\ref{iwamoto})
into
Eq. (\ref{dielectric})
are shown in Fig. 2.
They indicate that for $r_s\sim 10$ the enhancement is quite pronounced.
However, the reliability of both approaches can still be questioned.

\section{Non-zero external field}
Now let us address the situation when the Zeeman splitting and  SO coupling are present simultaneously. We will study the interplay between  the interaction-induced enhancement of the $g$ factor and of the SO coupling. First assume that Zeeman splitting is caused by a perpendicular magnetic field. The bare Hamiltonian in
this case takes the form
\begin{equation}
\label{hzso}
\hat{H}=E_{+}^{(0)}(k)\hat{{\cal P}}_{0\perp}^+({\bf k})+E_{-}^{(0)}(k)\hat{{\cal P}}_{0\perp}^-({\bf k}),
\end{equation}
where the modified projection operators
\begin{equation}
\label{modprojection}
\hat{{\cal P}}_{0\perp}^+({\bf k})=\frac{\gamma_0(k)}{1+\gamma_0^2(k)}\left(\begin{array}{cc}\gamma^{-1}_0(k) & ie^{-i\phi_{{\bf k}}}\\ -ie^{i\phi_{\bf k}} & \gamma_0(k)\end{array}\right),\hspace{1.5cm}\hat{{\cal P}}_{0\perp}^{-}({\bf k})=1-\hat{{\cal P}}_{0\perp}^{+}({\bf k}),
\end{equation}
are introduced. In Eq. (\ref{modprojection}) $\gamma_0(k)$ is defined as
\begin{equation}
\label{gammak}
\gamma_0(k)=\sqrt{1+\biggl(\frac{\Delta_Zk_F}{\Delta_{SO}k}\biggr)^2}-\frac{\Delta_Zk_F}{\Delta_{SO}k}.
\end{equation}
The bare energy spectrum is given by
\begin{equation}
\label{ekzso}
E_{\pm}^{(0)}(k)=\frac{\hbar^2k^2}{2m}\pm\frac{1}{2}\sqrt{\Delta^2_{SO}\biggl(\frac{k}{k_F}\biggr)^2+\Delta^2_Z},
\end{equation}
so that the splitting of the spectrum at $k=k_F$ equals
\begin{equation}
\label{delta}
\Delta=\sqrt{\Delta^2_{SO}+\Delta^2_{Z}}.
\end{equation}
The general expression Eq. (\ref{sigmak}) for the self-energy retains its form in the present case after changing $\hat{P}^{\pm}({\bf k})$ by $\hat{{\cal P}}_{\perp}^{\pm}({\bf k})$, where the renormalized projection operators have the form of Eq. (\ref{modprojection})
\begin{equation}
\label{modprojectionk}
\hat{{\cal P}}_{\perp}^+({\bf k})=\frac{\gamma(k)}{1+\gamma^2(k)}\left(\begin{array}{cc}\gamma^{-1}(k) & ie^{-i\phi_{{\bf k}}}\\ -ie^{i\phi_{\bf k}} & \gamma(k)\end{array}\right),\hspace{1.5cm}\hat{{\cal P}}_{\perp}^{-}({\bf k})=1-\hat{{\cal P}}_{\perp}^{+}({\bf k}),
\end{equation}
with renormalized parameter $\gamma(k)$, which should be determined self-consistently together with renormalized spectrum $ E_{\pm}(k)$. Since in the present case the operators $\hat{{\cal P}}_{\perp}^{\pm}({\bf k})$ differ from  $\hat{{\cal P}}_{0\perp}^{\pm}({\bf k})$, the consequence (\ref{form1}) of Eq. (\ref{sigmak}) is not valid anymore. Instead we get the following system of equations
\begin{eqnarray}
\label{eq1}
\frac{E_+(k)+\gamma^2(k)E_-(k)}{1+\gamma^2(k)}=\frac{E_+^{(0)}(k)+\gamma_0^2(k)E_-^{(0)}(k)}{1+\gamma_0^2(k)}+\\ \nonumber
\int\frac{d^2{\bf k}^{'}}{(2\pi)^2}\frac{V_{eff}(|{\bf k}-{\bf k}^{'}|)}{1+\gamma^2(k^{'})}\biggl[f_0(E_F-E_{+}(k^{'}))+\gamma^2(k^{'})f_0(E_F-E_{-}(k^{'}))\biggr],
\end{eqnarray}
\begin{eqnarray}
\label{eq2}
\frac{E_-(k)+\gamma^2(k)E_+(k)}{1+\gamma^2(k)}=\frac{E_-^{(0)}(k)+\gamma_0^2(k)E_+^{(0)}(k)}{1+\gamma_0^2(k)}+\\ \nonumber
\int\frac{d^2{\bf k}^{'}}{(2\pi)^2}\frac{V_{eff}(|{\bf k}-{\bf k}^{'}|)}{1+\gamma^2(k^{'})}\biggl[f_0(E_F-E_{-}(k^{'}))+\gamma^2(k^{'})f_0(E_F-E_{+}(k^{'}))\biggr],
\end{eqnarray}
\begin{eqnarray}
\label{eq3}
\frac{\gamma(k)}{1+\gamma^2(k)}\biggl[E_+(k)-E_-(k)\biggr]=\frac{\gamma_0(k)}{1+\gamma_0^2(k)}\biggl[E_+^{(0)}(k)-E_-^{(0)}(k)\biggr]+\\ \nonumber
+\int\frac{d^2{\bf k}^{'}}{(2\pi)^2}\cos(\phi_{\bf k}-\phi_{{\bf k}^{'}})V_{eff}(|{\bf k}-{\bf k}^{'}|)\biggl[f_0(E_F-E_{+}(k^{'}))-f_0(E_F-E_{-}(k^{'}))\biggr].
\end{eqnarray}

Subtracting Eq. (\ref{eq2}) from Eq. (\ref{eq1}), we get
\begin{eqnarray}
\label{eq4}
\frac{1-\gamma^2(k)}{1+\gamma^2(k)}\biggl[E_+(k)-E_-(k)\biggr]=\frac{1-\gamma_0^2(k)}{1+\gamma_0^2(k)}\biggl[E_+^{(0)}(k)-E_-^{(0)}(k)\biggr]+\\ \nonumber
+\int\frac{d^2{\bf k}^{'}}{(2\pi)^2}V_{eff}(|{\bf k}-{\bf k}^{'}|)\frac{1-\gamma^2(k^{'})}{1+\gamma^2(k^{'})}\biggl[f_0(E_F-E_{+}(k^{'}))-f_0(E_F-E_{-}(k^{'}))\biggr].
\end{eqnarray}
Now we can apply to Eq. (\ref{eq3}) and Eq. (\ref{eq4}) the same argument that led to renormalization of $\Delta_{SO}$ and $\Delta_{Z}$ respectively. In the zero-temperature limit this results in the following system of equations
\begin{equation}
\label{quadra1}
\frac{\gamma(k)}{1+\gamma^2(k)}\biggl[E_+(k)-E_-(k)\biggr]\biggl[1-\frac{m^*}{m}\lambda_{SO}\biggr]=\frac{\gamma_0(k)}{1+\gamma_0^2(k)}\biggl[E_+^{(0)}(k)-E_-^{(0)}(k)\biggr],
\end{equation}
\begin{equation}
\label{quadra2}
\frac{1-\gamma^2(k)}{1+\gamma^2(k)}\biggl[E_+(k)-E_-(k)\biggr]\biggl[1-\frac{m^*}{m}\lambda_{Z}\biggr]=\frac{1-\gamma_0^2(k)}{1+\gamma_0^2(k)}\biggl[E_+^{(0)}(k)-E_-^{(0)}(k)\biggr].
\end{equation}
Dividing Eq. (\ref{quadra2}) by Eq. (\ref{quadra1}) we get a closed quadratic equation for $\gamma(k)$
\begin{equation}
\label{qua}
\gamma^2(k)+\gamma(k)\frac{1-\gamma^2_0(k)}{\gamma_0(k)}\frac{1-\frac{m^*}{m}\lambda_{SO}}{1-\frac{m^*}{m}\lambda_{Z}}-1=0.
\end{equation}
Substituting the solution of this equation back into Eq. (\ref{eq1}),  we get the renormalized splitting of the spectrum $\Delta^*=E_+(k_F)-E_-(k_F)$
\begin{equation}
\label{delta*}
\frac{\Delta^*}{\Delta}=\frac{\sqrt{(1-\gamma^2_0)^2(1-\frac{m^*}{m}\lambda_{SO})^2+4\gamma_0^2(1-\frac{m^*}{m}\lambda_{Z})^2}}{(1+\gamma_0^2)(1-\frac{m^*}{m}\lambda_{SO})^2(1-\frac{m^*}{m}\lambda_{Z})^2}.
\end{equation}
Using the definition (\ref{gammak}) of $\gamma_0$, we can rewrite the last result in the following concise form
\begin{equation}
\label{triangle}
\frac{\Delta^*}{\Delta}=\sqrt{\frac{\Delta_Z^2}{\Delta_Z^2+\Delta_{SO}^2}\biggl(\frac{1}{1-\frac{m*}{m}\lambda_{Z}}\biggr)^2+\frac{\Delta_{SO}^2}{\Delta_Z^2+\Delta_{SO}^2}\biggl(\frac{1}{1-\frac{m*}{m}\lambda_{SO}}\biggr)^2}.
\end{equation}
Finally, with the use of Eqs. (\ref{effective}) and (\ref{deltaso}), we 
arrive to the conclusion that renormalized splitting $\Delta^*$ is related to renormalized values $\Delta^*_Z$ and $\Delta^*_{SO}$ in the same way as the bare values (Eq. (32))
\begin{equation}
\Delta^{*}=\sqrt{\Delta^{*2}_{SO}+\Delta^{*2}_{Z}}.
\end{equation}

Consider now the case when  the Zeeman splitting is caused by a parallel magnetic field applied along the $x$-direction. Then the Hamiltonian can be written as
\begin{equation}
\label{parallelfield}
\hat{H}=\left(\begin{array}{cc}\frac{\hbar^2k^2}{2m} & \frac{\Delta_Z}{2}+i\alpha k e^{-i\phi_{{\bf k}}}\\ \frac{\Delta_Z}{2}-i\alpha k e^{i\phi_{\bf k}} & \frac{\hbar^2k^2}{2m}\end{array}\right)=E_{+}^{(0)}(k)\hat{{\cal P}}_{0\parallel}^+({\bf k})+E_{-}^{(0)}(k)\hat{{\cal P}}_{0\parallel}^-({\bf k}),
\end{equation}
where the energy spectrum 
\begin{equation}
\label{parallelspectrum}
E_{\pm}^{(0)}(k)=\frac{\hbar^2k^2}{2m}\pm\frac{1}{2}\sqrt{\Delta_Z^2+\Delta^2_{SO}\biggl(\frac{k}{k_F}\biggr)^2+2\Delta_Z\Delta_{SO}\frac{k}{k_F}\sin\phi_{{\bf k}}},
\end{equation}
depends  both on the amplitude and orientation of ${\bf k}$ with respect to
the magnetic field. In Eq. (\ref{parallelfield}),  the projection operators $\hat{{\cal P}}_{0\parallel}^{\pm}({\bf k})$ are  defined as
\begin{equation}
\label{p+-para}
\hat{{\cal P}}_{0\parallel}^{+}({\bf k})=\frac{1}{2}\left(\begin{array}{cc}1 & ie^{-i\varphi_{{\bf k}}}\\ -ie^{i\varphi_{\bf k}} & 1\end{array}\right),\hspace{1.5cm}\hat{{\cal P}}_{0\parallel}^{-}({\bf k})=1-\hat{{\cal P}}_{0\parallel}^{+}({\bf k}),
\end{equation}
with the angle $\varphi_{{\bf k}}$  related to the azimuthal angle of vector ${\bf k}$ as follows
\begin{equation}
\label{varphik}
\varphi_{{\bf k}}=\arctan\biggl(\frac{\alpha k\cos\phi_{{\bf k}}}{\alpha k\sin\phi_{{\bf k}}+\frac{\Delta_Z}{2}}\biggr).
\end{equation}
The bare splitting of the energy spectrum at $|{\bf k}|=k_F$ is equal to 
\begin{equation}
\label{deltaparallel}
\Delta(\phi)=\sqrt{\Delta^2_Z  +\Delta^2_{SO}+2\Delta_Z\Delta_{SO}\sin\phi}.
\end{equation}
Performing calculations similar to those for perpendicular  field, it is easy to check that in the present case the relation between the renormalized splitting $\Delta^*$ and $\Delta_Z$,  $\Delta_{SO}$ preserves the form (\ref{deltaparallel})
\begin{equation}
\label{deltaphi}
\Delta^*(\phi)=\sqrt{\Delta^{*2}_Z +\Delta^{*2}_{SO}+2\Delta^{*}_Z\Delta^{*}_{SO}\sin\phi}.
\end{equation}

\section{Connection to the Landau Parameters}

The above calcualations were based on the concept of effective interaction between electrons, $V_{eff}(q)$. Generally speaking, Fermi liquid theory relates the observable values to the bare parameters of electron gas by means of interaction function \cite{pines,abr} having the form
\begin{equation}
\label{intfun}
f_{\sigma\sigma^{'}}({\bf k},{\bf k}^{'})=f^{s}({\bf k},{\bf k}^{'})+({\bf\bbox{ \sigma}}\cdot{\bf \bbox{\sigma}^{'}})f^{a}({\bf k},{\bf k}^{'})=\frac{\pi\hbar^2}{m^{*}}\biggl[F^{s}({\bf k},{\bf k}^{'})+({\bf\bbox{ \sigma}}\cdot{\bf \bbox{\sigma}^{'}})F^{a}({\bf k},{\bf k}^{'})\biggr],
\end{equation}
where ${\bf \bbox{\sigma}}$ and ${\bf \bbox{\sigma}^{'}}$ are spin matrices, $f^s({\bf k},{\bf k}^{'})$ and $f^a({\bf k},{\bf k}^{'})$ are the symmetric and antisymmetric parts  of the interaction function,  respectively. In Eq. (\ref{intfun}) $ F^{s(a)}({\bf k},{\bf k}^{'})=\sum_{l=0}^{\infty}F_l^{s(a)}\cos(l\phi_{{\bf k}{\bf k}^{'}})$ are dimensionless quantities. The concept of effective interaction used above  is equivalent to the assumption $f^a\equiv f^s$. The way to extend our theory in order to take into account the difference between $f^s$ and $f^a$ is to modify the self-energy Eq. (\ref{sigmak}) as follows
\begin{eqnarray}
\label{modsigma}
\hat{\Sigma}({\bf k})=\int\frac{d^2{\bf k}^{'}}{(2\pi)^2}\hat{P}^{+}({\bf k}^{'})\biggl[V_{eff}(|{\bf k}-{\bf k}^{'}|)f_0(E_F-E_{+}(k^{'})) \\ \nonumber
+W_{eff}(|{\bf k}-{\bf k}^{'}|)f_0(E_F-E_{-}(k^{'}))\biggr] \\ \nonumber
+\int\frac{d^2{\bf k}^{'}}{(2\pi)^2}\hat{P}^{-}({\bf k}^{'})\biggl[V_{eff}(|{\bf k}-{\bf k}^{'}|)f_0(E_F-E_{-}(k^{'})) \\ \nonumber
+W_{eff}(|{\bf k}-{\bf k}^{'}|)f_0(E_F-E_{+}(k^{'}))\biggr].
\end{eqnarray}
Here $V_{eff}(q)$ corresponds to the effective interaction in Eq. (\ref{sigmak}),  whereas  $W_{eff}(q)$ accounts for the difference between $f^s$ and $ f^a$. It is straightforward to check that with self-energy operator Eq. (\ref{modsigma}),  the projection operators $\hat{P}^{+}({\bf k})$ and $\hat{P}^{-}({\bf k})$ which,  in principle,  should be determined self-consistently, still retain the form Eq. (\ref{p+-}). Thus  we can repeat the derivation for the enhancement of $\Delta_{SO}$ in a similar way as  in Sec. II. The difference is,  however,  that the relation Eq. (\ref{real}) does not hold anymore. Indeed,  the renormalization of the effective mass\cite{kwon} is now determined by the effective interaction  $V_{eff}(q)+W_{eff}(q)$ through $\frac{m^{*}}{m}=1+\frac{1}{2}F^{s}_1$, while $\lambda_{SO}$ is  determined by  the first Fourier harmonics of $V_{eff}(q)-W_{eff}(q)$; accordingly,  $\lambda_{Z}$ is determined by the zero's Fourier component of  $V_{eff}(q)-W_{eff}(q)$. Consequently, in terms of the dimensionless  Landau parameters \cite{pines}, we get the following generalization of Eq. (\ref{simple})
\begin{equation}
\label{new}
\frac{\Delta_{SO}^{*}}{\Delta_{SO}}=\frac{1}{1+\frac{1}{2}F_1^{a}}.
\end{equation}
Note that,  while the dependence of $F_1^{s}$ on $r_s$ in two dimensions  was a subject of Monte Carlo studies \cite{tan,kwon}, the dependence $F_1^{a}(r_s)$ in 2D cannot be extracted form the current literature.

\section{Conclusion}

The main goal of this paper is to demonstrate that alongside with
fundamental characteristics, $\lambda_Z(r_s)$, of interacting electron gas,
which describes the  enhacement of the $g$-factor, and was
studied in many works, there exists another fundamental characteristics
$\lambda_{SO}(r_s)$ which describes the interaction-induced enhancement
of the SO coupling.
We calculated this function analytically in the limit of high concentrations
and estimated numerically at low concentrations. Note that throughout the
paper we assumed  the bare SO coupling to be small: $\Delta_{SO}\ll E_F$.
 However, in the limit
of high concentrations ($r_s\ll 1$) the corresponding condition is more strict:
$\Delta_{SO}\ll r_s E_F$, which is  equivalent to $\alpha \ll e^2/\varepsilon_0$.
In the intermediate region $r_sE_F \ll \Delta_{SO} \ll E_F$ the enhancement factor
is given by 
\begin{equation}
\label{intermed}
\lambda_{SO}=\frac{r_s}{\sqrt{2}\pi}\ln\Biggl(\frac{E_F}{\Delta_{SO}}\Biggr).
\end{equation}

Since experimentally the concentration of carriers is varied by changing
the gate voltage\cite{engels,nitta,shay}, there exists another simple reason 
for the dependence of the SO
coupling on the concentration. Indeed, the change of the gate voltage
causes the redistribution of the confining potential, which, in turn \cite{ross},
affects the parameter $\alpha$. This mechanism should be dominant
at high concentrations when $\lambda_{SO}$ is small.

Note in conclusion, that if the bare SO splitting is caused
by the Dresselhaus mechanism\cite{dress'}, which originates from the
absence of the inversion symmetry in the bulk, the renormalization of the
corresponding splitting of the spectrum, $\Delta_D$, has
the same form as Eq. (\ref{simple}): $\Delta_D^{\ast}=(1+\lambda_{SO})\Delta_D$.
As a result, when both $\Delta_D$ and $\Delta_{SO}$ are present, the
splitting of the energy spectrum is given by the same formula as for
noninteracting electrons\cite{erasm,pik}
\begin{equation}
\label{both}
\Delta^{\ast}(\phi)=
\sqrt{\Delta_D^{*2}+\Delta_{SO}^{*2}+2\Delta_D^{*}\Delta_{SO}^{*}\sin2\phi}.
\end{equation}

Finally, let us point out that in conventional magnetotransport 
oscillations experiments performed up to 
now\cite{luo',das1,engels,ram,nitta,heida} the typical concentrations
of electrons was quite high $\sim 10^{12}$cm$^{-2}$. As a result, the
typical values of the interaction parameter $r_s$ were rather low ($r_s < 1$).
Only in 2D hole gas\cite{shay} the condition ($r_s \geq 1$) was fulfilled.
For low values of $r_s$ our theory predicts that the renormalization
of the SO coupling is weak.
However, in recent experiments on the electron gases
 in silicon metal-oxide-semiconductor 
field-effect transistors\cite{kra,pud,pop} and AlAs quantum wells\cite{pap},
as well as in 
hole gases in SiGe quantum wells\cite{col}, 
GaAs inverted semiconductor-insulator-semiconductor structures\cite{han}, and
GaAs-AlGaAs heterostructures\cite{sim} the values of $r_s$ ranged from\cite{kra,pud,pop}
$r_s\approx 6$ to\cite{han,sim} $r_s\approx 24$. For such large  $r_s$
we predict a strong renormalization of the SO coupling, which might be of
relevance for metal-insulator transition  observed in these systems. For example, the strong renormalization of SO coupling at large $r_s$ might cause an instability of electronic spectrum in a clean system, so  that the system in zero magnetic field would undergo a transition into an exotic ``chiral phase'' at some  critical density. To be more specific, in the latest publication\cite{ha} the critical density for metal-insulator transition in n-type GaAs was reported to be $n_c\approx 1.3\times 10^{10}$ cm$^{-2}$. To accomodate all these electrons within the lower branch of the spectrum,  corresponding to chirality ``$-$'' (see Eq. (15)), the effective coupling constant $\alpha$ should exceed $\alpha_c=\biggl(\frac{\hbar^2n_c}{2\pi m^{*}}\biggr)^{1/2}\approx 3.65\times 10^{-12} $eV$\cdot$m. On the other hand, the constant $\alpha$ for relatively high-density GaAs/AlGaAs structure with $n\approx 4.0\times 10^{11}$cm$^{-2}$ can be extracted from Ref.\onlinecite{ram} to be $\Delta_{SO}/\sqrt{8\pi n}\approx 1.7\times 10^{-12}$eV$\cdot$m. Thus,  a two times interaction-induced enhancement of SO coupling (although Fig.2 for $r_s\approx 5$ shows a lesser value) would make a chiral phase feasible.
\vspace{0.5cm}
\begin{flushleft} \Large \bf
Acknowledgements
\end{flushleft}
We are grateful to R. R. Du, D. Mattis, and Y. S. Wu for useful discussions. We also thank L. E. Zhukov for assistance in preparation of the manuscript. M. E. R. ackowledges the discussions with H. U. Baranger and Y. Imry and the hospitality of the Aspen Center for Physics.

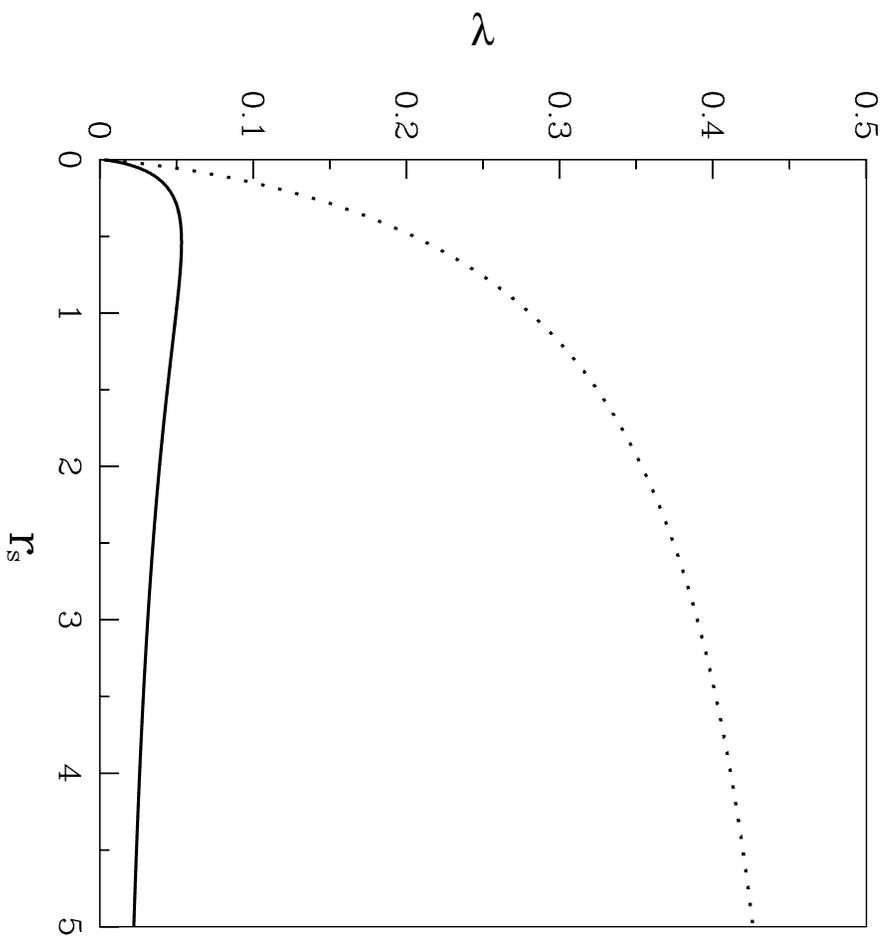
\begin{figure}
\caption{ The enhancement factors of Zeeman splitting (dotted line) and spin-orbit
splitting (full line), calculated within the static random phase approximation,
are plotted vs the interaction parameter $r_s$.}
\end {figure}
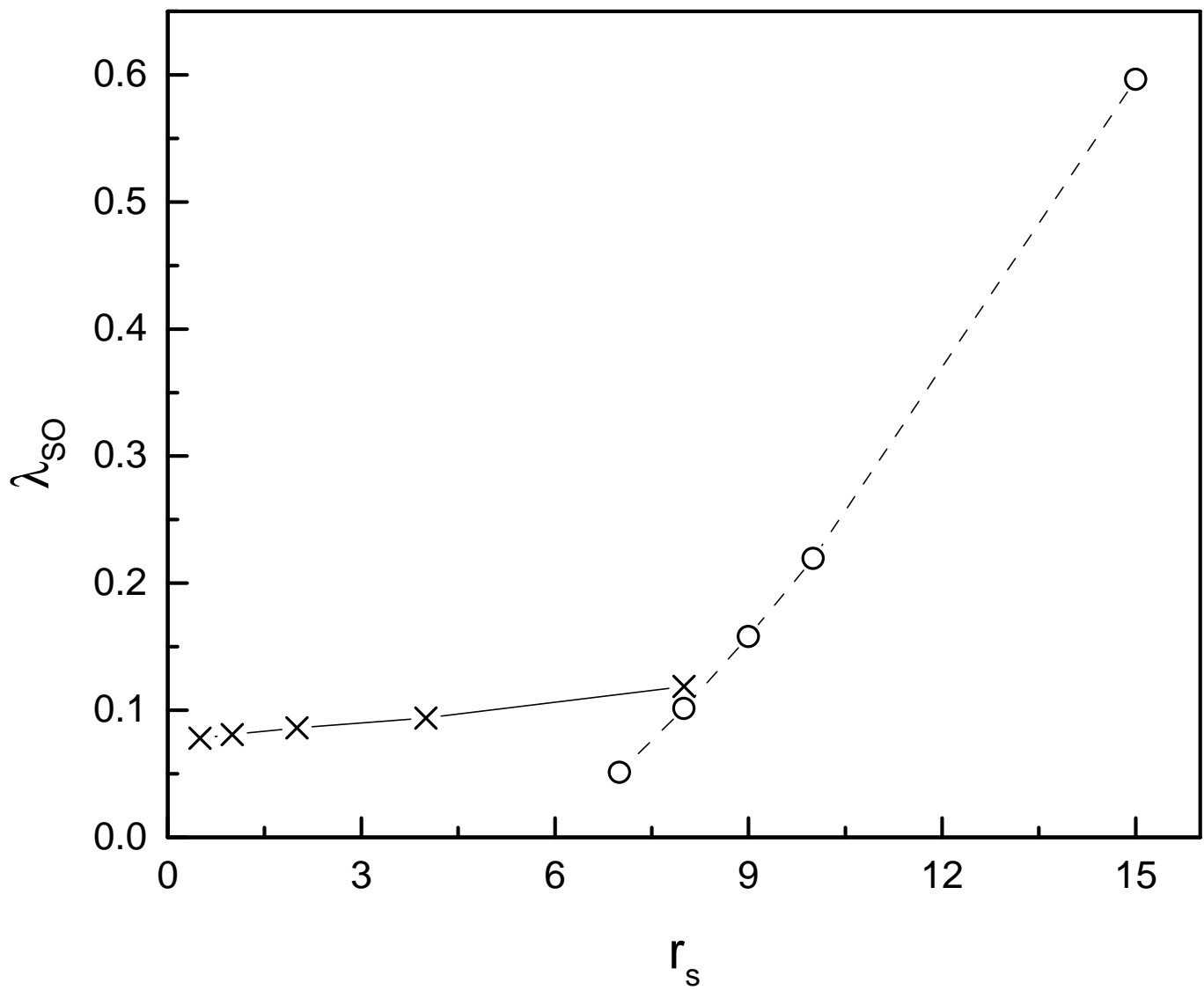
\begin{figure}
\caption{The enhancement factor $\lambda_{SO}$, calculated numerically with local field correction  taken into account, is plotted  vs the interaction parameter $r_s$. Full curve corresponds to the approach of Refs. 4 and 29 with  $G_{\infty}$ taken from Ref. 31 at points marked with crosses. Dashed curve is  calculated using the local field factor taken from Ref. 32  at points marked with  empty circles. }
\end{figure}


\vspace{0.5cm}
\begin{references}
\bibitem{fang}  F. F. Fang and P. J. Stiles, Phys. Rev. {\bf 174}, 823 (1968).
\bibitem{janak} J. F. Janak, Phys. Rev. {\bf 178}, 1416 (1969).

\bibitem{pines} D. Pines and P. Nozi\'{e}res, {\em The Theory of Quantum Liquids}
 (Benjamin, Reading, Mass., 1966).
\bibitem{yarl} S. Yarlagadda and G. F. Giuliani,
 Phys. Rev. B {\bf 40}, 5432 (1989).

\bibitem{stern} F. Stern, {Phys. Rev. Lett.}, {\bf 18}, 546(1967).

\bibitem{ale}I. L. Aleiner and L. I. Glazman, Phys. Rev. B {\bf 52},
11296 (1995).



\bibitem{lee}C. S. Ting, T. K. Lee, and J. J. Quinn, Phys. Rev. Lett. {\bf 34},
870 (1975); T. K. Lee, C. S. Ting, and J. J. Quinn, {\em ibid}. {\bf 35}, 1048
(1975).

\bibitem{kwon}Y. Kwon, D. M. Ceperley, and R. M. Martin, Phys. Rev. B {\bf 50},
1684 (1994).


\bibitem{luo}J. Luo, H. Munekata, F. F. Fang, and P. J. Stiles,
Phys. Rev. B {\bf 38}, 10142 (1988). 

\bibitem{das} B. Das, D. C. Miller, S. Datta, R. Reifenberger, W. P. Hong,
P. K. Bhattacharya, J. Singh, and M. Jaffe, Phys. Rev. B {\bf 39}, 1411 (1989).
 
\bibitem{rashba}  Yu. A. Bychkov and E. I. Rashba, Pis'ma Zh. Eksp. Teor. Fiz.
 {\bf 39}, 64 (1984) [JETP Lett. {\bf 39}, 78 (1984)].
\bibitem{rashba'} Yu. A. Bychkov and E. I. Rashba, J. Phys. C {\bf 17},
6039 (1984).
\bibitem{klitz}D. Stein, K. v. Klitzing, and G. Weinmann, Phys. Rev. Lett. {\bf 51}, 130 (1983).
\bibitem{stor}H. L. Stormer, Z. Schlesinger, A. Chang, D. C. Tsui,
A. C. Gossard, and W. Wiegmann, Phys. Rev. Lett. {\bf 51}, 126 (1983).



\bibitem{luo'}J. Luo, H. Munekata, F. F. Fang, and P. J. Stiles,
Phys. Rev. B  {\bf 41}, 7685 (1990).

\bibitem{rash''}Yu. A. Bychkov, V. I. Mel'nikov, and E. I. Rashba,
 Zh. \'{E}ksp. Teor. Fiz. {\bf 98}, 717 (1990), 
[Sov. Phys. JETP {\bf 71}, 401 (1990)].

\bibitem{das1}B. Das, S. Datta, and R. Reifenberger, Phys. Rev. B {\bf 41},
8278 (1990).


\bibitem{engels}G. Engels, J. Lange, Th. Sch\"{a}pers, and H. L\"{u}th,
Phys. Rev. B {\bf 55}, R1958 (1997).

\bibitem{ram}P. Ramvall, B. Kowalskii, and P. Omling, Phys. Rev. B {\bf 55}, 7160
(1997).

\bibitem{nitta} J. Nitta, T. Akazaki and H. Takayanagi, Phys. Rev. Lett. {\bf 78}, 1335 (1997).

\bibitem{heida} J. P. Heida, B. J. van Wees, J. J. Kuipers,  and T. M. Klapwijk, {Phys. Rev. B}, {\bf 57}, 11911 (1998).

\bibitem{shay} J. P. Lu, J. B. Yau, S. P. Shukla, M. Shayegan,
L. Wissinger, U. R\"{o}ssler, and R. Winkler, Phys. Rev. Lett. {\bf 81},
1282 (1998).





\bibitem{dress}P. D. Dresselhaus, C. M. A. Papavassilou, R. G. Wheeler,
and R. N. Sacks, Phys. Rev. Lett. {\bf 68}, 106 (1992).


\bibitem{mathur}H. Mathur and A. D. Stone, Phys. Rev. Lett. {\bf 68},
2964 (1992).



\bibitem{lyanda}see the most recent publication Y. Lyanda-Geller, 
Phys. Rev. Lett. {\bf 80}, 4273 (1998), and references therein.
The effect of the SO term on magnetotransport in the
insulating regime was  studied theoretically in T. V. Shahbazyan
and M. E. Raikh, Phys. Rev. Lett. {\bf 73}, 1408 (1994). 

\bibitem{abr}A. A. Abrikosov, L. P. Gor'kov, and I. E. Dzyaloshinski,
{\em Methods of Quantum Field Theory in Statistical Physics} (Pergamon, New York,
 1965).




\bibitem{mahan} G. D. Mahan, {\em Many Particle physics} (Plenum, New York, 1990).

\bibitem{yar} S. Yarlagadda and G. F. Giuliani,  Phys. Rev. B {\bf 38}, 
10966 (1988).
\bibitem{san}G. E. Santoro and G. F. Giuliani, Solid. State Commun. {\bf 67},
681 (1988).
\bibitem{yarlag} S. Yarlagadda and G. F. Giuliani, 
Solid. State Commun. {\bf 69},
677 (1989).
\bibitem{jonson}M. Jonson, J. Phys. C {\bf 9}, 3055 (1976).


\bibitem{iwam}  N. Iwamoto, {Phys. Rev. B}, {\bf 43}, 2174 (1991).

\bibitem{tan}B. Tanatar and D. M. Ceperley, Phys. Rev. B {\bf 39}, 5005 (1989).


\bibitem{ross}U. R\"{o}ssler, L. Malcher, and G. Lommer, Phys. Rev. Lett.
{\bf 60}, 728 (1988).


\bibitem{dress'} G. Dresselhaus, Phys. Rev., {\bf 100}, 580 (1955).




\bibitem{erasm}E. A de Andrada e Silva, Phys. Rev. B {\bf 46}, 1921 (1992).

\bibitem{pik}F. G. Pikus and G. E. Pikus, Phys. Rev. B {\bf 51}, 16928 (1995).






\bibitem{kra}  S.V. Kravchenko, D. Simonian, M. P. Sarachik, W. Mason,  and J. E. Furneaux,  Phys. Rev. Lett. {\bf 77}, 4938 (1996).

\bibitem{pud}V. M. Pudalov,  Pis'ma Zh. Eksp. Teor. Fiz. {\bf 66}, 168 (1997) 
[JETP Lett. {\bf 66}, 175 (1997)].

\bibitem{pop}D. Popovi\'{c}, A. B. Fowler, 
and S. Washburn, Phys. Rev. Lett. {\bf 79},
1543 (1997).


\bibitem{pap} S. J. Papadakis and M. Shayegan, Phys. Rev. B {\bf 57},
R15068 (1998).



\bibitem{col}P. T. Coleridge, R. L. Williams, Y. Feng, and P. Zawadzki, Phys. Rev. B {\bf 56}, R12764 (1997).

\bibitem{han}  Y. Hanien, U. Meirav, D. Shahar, C. C. Li, D. C. Tsui,  and H. Shtrikman,   Phys. Rev. Lett. {\bf 80}, 1288 (1998).

\bibitem{sim} M. Y. Simons, A. R. Hamilton, M. Pepper,
 E. H. Linfield, P. D. Rose, D. A. Ritchie, A. K. Savchenko, and T. G. Griffiths,  Phys. Rev. Lett. {\bf 80}, 1292 (1998). 
\bibitem{ha}  Y. Hanien, D. Shahar, J. Yoon,  C. C. Li, D. C. Tsui,  and H. Shtrikman,   Phys. Rev. B. {\bf 58}, R13338 (1998).

\end{references}
\end{document}